\documentclass[prl,showpacs, twocolumn,superscriptaddress]{revtex4}
\usepackage{graphicx}
\begin{document}

\title{Local environment of Nitrogen in GaN$_{y}$As$_{1-y}$ epilayers on GaAs (001) studied
using X-ray absorption near edge spectroscopy}

\author{J. A. Gupta[*]}
\author{M. W. C. Dharma-wardana}
\affiliation{
Institute for Microstructural Sciences, National Research Council of Canada,
Ottawa, Canada K1A 0R6} 
\author{A. J\"{u}rgensen}
\affiliation{Canadian Synchrotron Radiation Facility, Madison WI, 53589-3097 USA}
\author{E.D. Crozier}
\affiliation{Dept of Physics, Simon Fraser University, Burnaby, Canada
V5A 1S6}
\author{J.J. Rehr}
\author{M.Prange}
\affiliation{Dept of Physics, University of Washington, Seattle, WA 98195-1560 USA} 

\date{\today}
\begin{abstract}
X-ray absorption near-edge spectroscopy (XANES) is used to study the
N environment in bulk GaN and in  GaN$_{y}$As$_{1-y}$ epilayers 
on GaAs (001), for $y\sim$5\%. 
Density-functional optimized structures were used to
predict XANES via multiple-scattering theory.
We obtain striking agreement for pure GaN.
An alloy model with  {\it nitrogen
 pairs} on Ga accurately predicts the threshold energy,
the width of the XANES ``white line'',
 and features above threshold, for the given 
 X-ray polarization. The presence of N-pairs may 
point to a role for molecular N$_2$ in epitaxial growth kinetics.

\end{abstract}

\maketitle

Nitrogen-alloyed GaAs, i.e., GaN$_{y}$As$_{1-y}$, is intensely studied owing to its technological
importance. 
Even very small amounts of N lead to a dramatic
$y$-dependent decrease in the bandgap\cite{wei}. 
In the ultradilute regime ($y<0.01$\%), 
 localized, single-impurity levels produce
sharp resonances above the GaAs conduction band minimum.
With increasing $y$, small nitrogen aggregates (pairs
and higher clusters) give rise to localized states
 in the band gap, producing
sharp photoluminescence (PL) lines~\cite{francoeur1538, liu7504}.
At higher concentrations ($y>0.1-0.25$\%), the sharp PL lines
 merge to a broader line.
Anomalous bandgap bowing and other effects are observed~\cite{kent115208, francoeur1538}
in this regime.  
At these concentrations, a 
phenomenological band anti-crossing model works well~\cite{shan,oreilly}, since here the
details of
N-incorporation seem to be unimportant.  
Kent and Zunger\cite{kent115208} presented a theoretical study using 
pseudopotentials and
large supercells containing various N-atom configurations.
In reality, a plethora of different local atomic environments is possible,
since Ga can have 0 to 4 first neighbour
N atoms; N pairs can have different orientations or 
separations; and triplets or higher order clusters are
possible~\cite{kent115208,kent2339}.
The change in PL on annealing has even been attributed to the
removal of N interstitials or defect complexes, such as N-As split interstitials
\cite{krispin2120,zhang1789}.

Several groups have used XANES
 to study local atomic structures in GaNAs epilayers.
Lordi {\it et al.}\cite{lordi}
 modeled their N XANES data using local densities of states (LDOS)
obtained from the Vienna {\it Ab Initio}
 Simulation Package(VASP)~\cite{kress}.
For GaNAs the partial LDOS did not reproduce the experimental  "white line"
and predicted several unobserved oscillatory features above the absorption edge.
Simpler tight-binding (TB) methods have also been examined by other authors~\cite{nodwell}.
In general TB is unreliable except for band-structure type calculations\cite{tit} to which the TB
parameters have been fitted.

In this Letter we present a study of N incorporation in GaN$_{y}$As$_{1-y}$ epilayers. 
Our objectives are: (i) to  compare  the local nitrogen environments
 in the dilute nitride and
in bulk GaN;
(ii) to 
 predict the
XANES spectra for several local atomic arrangements of N in GaN$_{y}$As$_{1-y}$ and
compare them with
experiment. These calculations used the {\it ab-initio} code FEFF8~\cite{ankudinov}
which combines simultaneous real-space full multiple scattering (FMS) calculations
of near-edge X-ray absorption spectra within a self-consistent field (SCF) treatment
of the electronic structure. 
The microscopic structure of the alloy epilayers for N or (N,N) pair-configurations
 were separately  determined via density-functional theory (DFT) using the
VASP code\cite{kress} within an ultrasoft pseudopotential scheme, and the 
generalized gradient approximation for the exchange-correlation potential.
The DFT optimized structure was used as the input to the XANES calculations.
The predicted XANES spectra
are in very close agreement with the experimental results, and support 
the presence of (N,N) pair clusters in this alloy regime ($y\sim 3-6$\%).

The single, uncapped GaN$_{y}$As$_{1-y}$ epilayer used
 here was grown by solid-source
MBE as described elsewhere\cite{guptambe}.
 The actual composition, $y=0.0496$, and thickness, 333~\AA, of the sample
were obtained from high-resolution X-ray diffraction data
via dynamical diffraction analysis. 
A thick, metalorganic vapor phase epitaxy-grown GaN layer on
 sapphire (0001) was used as a reference.
Experimental XANES measurements were performed at the
 Canadian Spherical Grating Monochromator 
beamline of the Synchrotron Radiation Center at the University of Wisconsin, Madison. 
The storage ring energy was 1~GeV, and a 600~lines/mm grating was used.
The measurements on GaN and GaNAs
epilayers used entrance-slit widths of 100~$\mu$m and 250~$\mu$m, 
providing energy resolution of $\sim$ 0.24~eV and 0.54~eV, respectively.
  Measurements at the nitrogen K-edge
were performed using both total electron yield (TEY)
and total fluorescence yield (TFY) modes of detection.
TEY provides more surface-sensitive information.
For GaN, the TEY and TFY data were nearly identical,
 but the TEY data exhibited a more intense main peak. 
For GaNAs, only the TFY data are presented here.
The samples were outgassed by {\it in-situ}
 heating to 280~$^{\circ}$C for $\sim$4 hours
in the ultra-high vacuum analysis chamber.
The effectiveness of this surface preparation was evident
 from the elimination of the Oxygen K-edge peak in the
absorption spectra.

Five GaN$_{y}$As$_{1-y}$ models were considered in this study.
For each structure, state-of-the-art 
 DFT calculations
within the  generalized gradient approximation (GGA) were carried out with
a 64 atom $2\times 2\times 2$
 supercell of zincblende unit cells, with periodic boundary conditions. 
Tetragonal distortions were included by holding the in-plane epilayer
lattice constant at the GaAs(substrate) value. The lattice
constant along the growth direction and atomic positions were
 relaxed until the Hellman-Feynman
forces became negligible, resulting in the lattice constant along the growth direction
being smaller than the in-plane value. 
 Each FEFF model was formed by ``stacking'' 27 relaxed DFT supercells 
 to form a  1792-atom, cubic supercell with the absorbing-N atom at the center.
  Spherical
clusters using only the 295 central atoms were used for the FEFF calculations.
A single N atom in the DFT cell yields 
the alloy concentration  $y=1/32$.
The artificial periodicity of the DFT supercell
resulted in second N atoms at
$\sim11.2$~\AA $\,$from the central, ``quasi-isolated"  N
absorber. All our models contain this artificial periodicity.
 Using the notation of Ref.~\onlinecite{kent115208},
 we also considered
(N,N) pairs in $m$th nearest neighbour positions
 with $m$=1, 2 and 6, corresponding to FCC lattice
positions (1/2 0 1/2), (1 0 0) and (\={1} 1 \={1}).
 For the $m$=2 case, we studied (N,N) pairs parallel 
($m$=2a) or perpendicular ($m$=2b, (0 0 \={1})) to the substrate, with the X-ray absorbing atom
at (0 0 0).
Thus, we have one ``single N'' model,
and four (N,N) pair models  labeled as NN$m$, i.e.,
\begin{equation}
\label{list}
\mbox{Single N, NN1, NN2a, NN2b and NN6}
\end{equation}  

For nitrogen K-edge XANES, large FEFF clusters
are needed to capture multiple scattering resulting from the high density
and the long mean free path resulting from the very long core-hole lifetime
(inverse lifetime = 0.116 eV).
In this work, 27 and 295-atom SCF and FMS clusters
 were used and parallel-processed on
32 nodes.

\begin{figure}[tbp]
\centering
\includegraphics*[width=8.0cm]{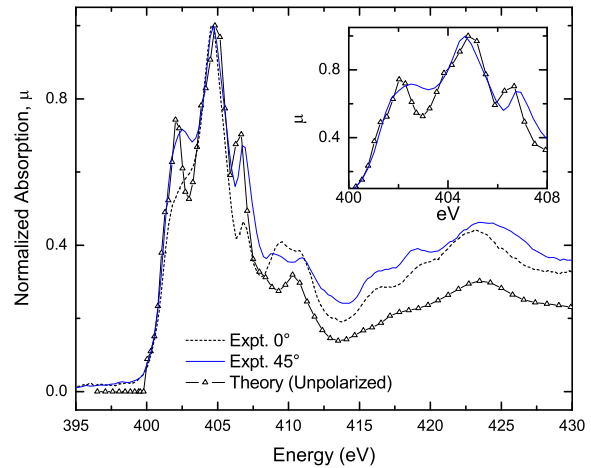}
\caption{Nitrogen K-edge XANES for wurtzite GaN compared with 
theory (unpolarized calculation). Incidence angles of 0$^{\circ}$
and 45$^{\circ}$ were used in single scans at a rate of 10~sec/point,
 and step size of 0.25~eV. The inset shows the excellent agreement
for incidence at 45$^{\circ}$, close to the Lambert angle.} 
\label{fig:ganxanes}
\end{figure}

Fig.~\ref{fig:ganxanes} shows the experimental data for the thick GaN reference epilayer
obtained at incidence angles of 0$^{\circ}$ and 45$^{\circ}$,
 relative to the surface normal.
The data show a rich fine structure consistent with previous
 reports~\cite{lambrecht,katsikini}.
These two data sets illustrate the importance of polarization effects,
(cf. Ref.~\onlinecite{lambrecht}).  
Most notably, the peaks near 402.5~eV and 406.8~eV are
 significantly enhanced at 45$^{\circ}$ compared to 
normal incidence. 
 The former is close to the ``magic'' angle,
 54.7$^{\circ}$ (the Lambert angle), where the
polarization effects would be completely suppressed~\cite{katsikini}.

The figure also shows the FEFF8 results for the GaN N~K-edge
XANES based on the hexagonal Wurtzite structure with the unit cell
 parameters $a$ = 3.186~\AA, $c$ = 5.178~\AA.
Clearly, the theory is very successful
 in predicting the main XANES features, especially at
45$^{\circ}$.
In particular, the
 edge energy is accurately predicted without any
 ``fitting'' to experiment.
The experimental absorption edge energy, $E_{0}=401.1\pm0.1$~eV
was determined from the zero of the second derivative
 of the absorption,
 while the value
$E_{0}=400.93$~eV was predicted by FEFF.  The experimental uncertainty
is about half the energy step (0.25eV).
On normalizing to match the height of the main peak, $\sim$405 eV,
 the theory accurately predicts the magnitudes
and positions of the peaks near 402.5, 406.8 and 411~eV,
 as well as the local minimum near 414 eV, the local maximum near 423 eV,
and the general decay of the absorption with increasing
photon energy.
A determination of the experimental
polarization was difficult given
 the irregular shape of the
sample after cleaving the sapphire substrate.
Hence the appropriate comparison is between
 our unpolarized model and the experimental data at 45$^{\circ}$.
Unlike the theoretical results of Ref.~\onlinecite{lambrecht},
our results show no overestimate of the peak positions at higher energies. 
This suggests that the multiple-scattering methods of FEFF8 can be used
for GaN$_{y}$As$_{1-y}$ with confidence,
if a realistic alloy structure is available.

\begin{figure}[tbp]
\centering
\includegraphics*[width=8.0cm, height=10.0cm]{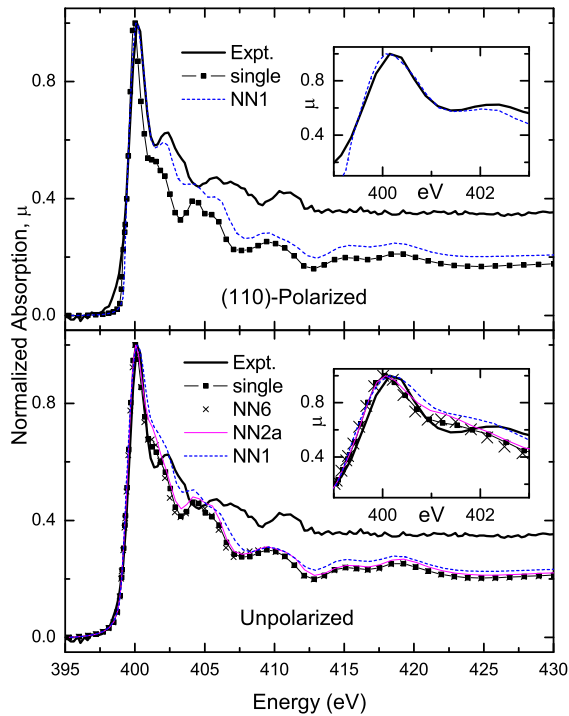}
\caption{Nitrogen K-edge XANES for a coherently-strained GaNAs epilayer
 (thick line) compared with theoretical models
listed in Eq.~\ref{list}.
The data
are single scans, with 135~sec/point, and a step size of 0.25~eV.
Top: for photons polarized along (110). 
Bottom: unpolarized photons. 
The XANES for NN2b is very similar to 
NN2a, and is omitted. The theoretical absorption asymptote for 
higher eV  is lower than in experiment as
self-absorption effects are not included in the theory.
} 
\label{fig:ganasxanes}
\end{figure}

Fig.~\ref{fig:ganasxanes} shows the experimental
 XANES for the GaN$_{y}$As$_{1-y}$ epilayer.
The local environment of the Nitrogen
in the alloy is clearly different from the pure GaN.  The absorption threshold in this
case has a single, narrow ``white line'', as opposed to the multiple peaks in GaN.
 A smaller, distinct
peak is observed at 402.3~eV, coincident with the lowest energy peak in pure GaN.
The top panel shows that the NN1 model accurately predicts the first and second peaks
when the photon polarization is included in the theory. The lower panel shows that the
unpolarized calculation may have some relevance to the threshold edge,
and for energies above $\sim 405$ eV.
The threshold energy was found to be
$E_{0}=399.6\pm 0.1$eV.
Our experimental data are consistent with those in
Refs.~\onlinecite{lordi} and~\onlinecite{nodwell}. 
 The data of Ref.~\onlinecite{strocov} are similar, although their second peak
intensity is almost at the ``white line".  

Several properties determined from the DFT 
 calculations are summarized in Table~\ref{table:modelsummary}.
The DFT predicted tetragonal distortions are in agreement with Vegard's law,
confirming its applicability in this alloy regime.

\begin{table}
\caption{Growth direction (001) lattice constants $a_{\perp}$, and
 average Ga-N bond lengths $b$ with standard deviation $\sigma_{b}$,
 and nearest nitrogen
distance $R$ obtained by total energy minimization for the alloy models
(see Eq.~\ref{list}  in the text) studied here.
N.B., the Ga-N and Ga-As bond lengths in cubic GaN
and GaAs are 1.9500 and 2.4480~\AA, respectively. The in-plane
lattice constant is constrained to the GaAs value, 5.6534~\AA.
}
\label{table:modelsummary}
\begin{ruledtabular}
\begin{tabular}{ccccc}
{\bf Model} & {\bf $a_{\perp}$ (\AA)} & {\bf $b$ (\AA) } &
 {\bf $\sigma_{b}$ (\AA)} &
 {\bf $R$ (\AA)}   \\ 
\hline \\
Single N &       5.5865  &       2.0567  & 0.0002 & 
  11.1730    \\ 
NN1    &       5.5131  &       2.0722  & 0.0416 & 
  3.6908 \\ 
NN2a   &       5.5108  &       2.0504  & 0.0010 & 
 5.6530  \\ 
NN2b   &       5.5195  &       2.0510  & 0.0007 & 
  5.5191 \\ 
NN6    &       5.5087  &       2.0468  & 0.0010 &
  9.7066 \\ 
\end{tabular}
\end{ruledtabular}
\end{table}
For the single-N atom model, the Ga-N bond lengths are nearly identical,
 $R=2.0567$\AA, while the
NN1 model predicts a distribution of bond lengths, $R=2.072\pm 0.042$\AA.
This is an indication of lattice distortions produced by this configuration,
and we find that this (N,N) pair configuration is not as energy optimal~\cite{kent115208}
as some other models. However, Molecular Beam Epitaxy is a non-equilibrium technique
hence the MBE-grown epilayer need not have the most energy-optimal N-cluster. The objective of the XANES is in fact to find
the most prevalent N cluster in the alloy sample prepared via MBE. 

Fig.~\ref{fig:ganasxanes}, bottom panel illustrates the spectral changes due to the local
N environment in the singlet and pair models, not including polarization effects.  
In this figure the FEFF8 spectra were shifted (by an amount smaller than the
experimental stepsize 0.25eV) to match the
experimental edge energy. 
The threshold energy $E_{th}$ is almost identical for
the different models, and hence independent of the
local bonding configurations and the number of N atoms
in these models.
The shift in the $E_{th}$ 
 between GaN and GaN$_{y}$As$_{1-y}$ (-1.45~eV) is thus a general feature
of N alloying in GaAs, and is related to long-range effects and the
 larger lattice constant of the GaN$_{y}$As$_{1-y}$
epilayer compared with GaN. 
This is consistent with the concept of nitrogen impurities
 forming ``perturbed host states'' in
GaAs~\cite{kent115208}.

The apparent discrepancy between the calculated intensities
 at higher energies compared with the experimental results is due to
a reduced white line intensity in our fluorescence data due to self-absorption effects;
our normalization using this lower peak intensity causes the apparent
 increase in the absorption at higher energies.
Otherwise, the ``white line" is reproduced extremely well by FEFF,
and is not affected by the local nitrogen
environment. The XANES white line is often understood 
 as arising from a high density of
final states or exciton effects due to the Coulomb interaction between the photoelectron
 and the core hole ~\cite{brown}.  
Nodwell {\it et al.} suggested that the white line is due to a N-related resonance
 in the conduction band and not to an excitonic
bound state~\cite{nodwell}.
 However, core hole effects were included in our FEFF calculations and provide
 an excellent description
of the data. 

The NN6 model and the single-N atom model have very similar spectra
(Fig.~\ref{fig:ganasxanes}, bottom panel),
 indicating that the phase interference between
the two atoms with the large NN6 separation is sufficiently small
 as to have no effect on the XANES, in agreement with the
negligible interaction energy for $m=6$ pairs
(see also Ref.~\onlinecite{kent115208}).
In the models where the second N atom is moved closer to the N absorber,
 the interaction as well as interference of scattered amplitudes
becomes more significant.
For the NN2 models
interactions produce a slight increase in the intensity of the
high-energy shoulder near the white line, but the rest of the NN2a and NN2b
 spectra are similar to the single and NN6 models.
A more dramatic effect is observed for the NN1 model.
 The intensity of the shoulder is increased and the
shoulder is shifted closer to the small peak at 402.3~eV of the experimental data.
All models predict a minimum near 413~eV, well-aligned
 with the falling edge of the final experimental peak.
The predicted features between 413~eV and about 419~eV are common
 for all models and similar to those in GaN 
in this energy range.  These spectral
 features are due mostly to Ga-N bonds,
 traditionally studied by EXAFS.

We return to the polarized photon calculation of Fig.~\ref{fig:ganasxanes}.
The absorption cross-section depends on the components of atomic bonds
 along the direction of the electric field
polarization.
 For the NN1 pair model, the NN1
neighbour is located at
(2.6438\AA, 0.000\AA, 2.5753\AA), so the (110) and ($\bar{1}$10) components are approximately equal. 
The agreement between theory and experiment is strong evidence that
NN1 pairs are responsible for the peak at 402.3~eV in the experimental data. 
We also note that the 402.3 eV peak is not shown by the other models
and is not well-resolved in several triplet models(see Ref.~ \onlinecite{kent115208}) that we examined,
 but have not reported in detail in
this paper. If the tetragonal distortion (coherent epilayers) arising from fixing the
in-plane lattice constant to the subtrate is relaxed, the resulting ``annealed''
NN1 model loses the 401.4 eV minimum (fig.2 top panel shows only the
coherent epilayer results) and the 402.3 eV peak moves to
lower energies and becomes a shoulder.
This type of peak narrowing is   
expected since the annealing reduces the bond-length distribution. In fact, the
width of the 402.3 eV peak in the experimental data (fig.2, top)
 as compared to the theory suggests that
the samples have more disorder than in the optimal tetragonal NN1 model used in the theory. 

Nodwell {\it et al.} noted that their XANES data were the same for samples
 with different N compositions~\cite{nodwell}.
This seems reasonable since 
 the observed fine structure contains major contributions from microscopic
N aggregates, such as nitrogen first-neighbour FCC pairs,
 and these structures are formed even
at very low N content.  
Ref.~\onlinecite{kent115208} considered several triplet structures, including
the Cs symmetry triplet proposed by Gil and Mariette~\cite{gil},
 which was found to be consistent with GaNP data.
The cluster
 formation does not result solely from the natural, statistical distribution of N atoms.
The low miscibility of N in GaAs has been suggested as a cause of 
locally ordered structures~\cite{neugebauer,guptathermo}.
An interesting possibility for the origin of close N,N pairs
is that they are favoured by the kinetics of the epitaxial growth
process. In fact, it has recently been proposed that molecular nitrogen
in an excited state is a dominant agent in GaN MBE.\cite{molecN}

In conclusion, high-resolution XANES of a
 coherently-strained GaN$_{y}$As$_{1-y}$ epilayer on GaAs(001) and a 
bulk GaN film were presented.  For pure GaN, our theoretical
 spectrum agrees extremely
well with the measured spectrum and demonstrates the effectiveness
 of the full multiple scattering approach within  
the self-consistent field calculations of the Fermi energy,
 and the importance of the synchrotron polarization.
For the strained GaN$_{y}$As$_{1-y}$ epilayer on GaAs(001),
 the calculated spectra exhibited a strong sensitivity to
the local N environment and we show that nitrogen
 FCC {\it pairs} attached to Ga captures the observed XANES.

{\it Acknowledgements}- We thank Z.R. Wasilewski for useful comments,
D. Ritchie for computer-system help, M. Tomlinson and S. Moisa for technical support,
and H. Tang for the GaN reference sample. CSRF is funded by
the National Research Council of Canada and an MFA grant from the Natural Sciences
and Engineering Research Council of Canada (NSERC). SRC is funded by the U.S. National
Science Foundation under grant No. DMR-0084402. One of us (EDC) acknowledges funding
from an NSERC Discovery Grant.


\begin{thebibliography}{99}

\bibitem{wei} S.H. Wei and A. Zunger, Phys. Rev. Lett. {\bf 76}, 664 (1996).

\bibitem{francoeur1538} S. Francoeur et al., Appl. Phys. Lett {\bf 75}, 1538 (1999).

\bibitem{liu7504} X. Liu, M.-E. Pistol and L. Samuelson, Phys. Rev. B {\bf 42}, 7504 (1990).

\bibitem{kent115208} P.R.C. Kent
and A. Zunger,
 Phys. Rev. B {\bf 64}, 115208 (2001).

\bibitem{shan} W. Shan et al., Phys. Rev. Lett. {\bf 82}, 1221 (1999).

\bibitem{oreilly} E.P. O'Reilly and A. Lindsay, Phys. Stat. Sol. B {\bf 215}, 131 (1999).

\bibitem{kent2339} P.R.C. Kent et al.,
 Appl. Phys. Lett. {\bf 79}, 2339 (2001).
\bibitem{krispin2120} P. Krispin et al., Appl. Phys. Lett, {\bf 80},
2120 (2002).

\bibitem{zhang1789} S.B. Zhang and S.-H. Wei, Phys. Rev. Lett. {\bf 86}, 1789 (2001).

\bibitem{lordi} V. Lordi et al., Phys. Rev. Lett. {\bf 90}, 145505 (2003).
\bibitem{kress}G. Kress, J. Furthmuller and J. Hafner,
see  http://cms.mpi.univie.ac.at/vasp/

\bibitem{strocov} V.N. Strocov et al., Phys. Stat. Sol. (b) {\bf 233}, R1 (2002).

\bibitem{nodwell} E. Nodwell et al., Phys. Rev. B {\bf 69}, 155210 (2004).
\bibitem{tit}N. Tit and M. W. C. Dharma-wardana, App. Phys. lett. {\bf 76}, 3576 (2000)

\bibitem{ankudinov} A.L. Ankudinov, B. Ravel, J.J. Rehr and S.D. Conradson, Phys. Rev. B {\bf 58}, 7565 (1998).

\bibitem{guptambe} J.A. Gupta et al., J. Cryst. Growth {\bf 242}, 141 (2002).

\bibitem{lambrecht} W.R.L. Lambrecht {\it et al.} Phys. Rev. B {\bf 55}, 2612 (1997).

\bibitem{katsikini} M. Katsikini et al., Phys. Rev. B {\bf 56}, 13380 (1997).

\bibitem{brown} M. Brown et al., Phys. Rev. B {\bf 15}, 738 (1977).

\bibitem{gil} B. Gil and H. Mariette, Phys. Rev. B {\bf 35}, 7999 (1987).

\bibitem{neugebauer} J. Neugebauer et al., Phy. Rev. B {\bf 51}, 10568 (1995).

\bibitem{guptathermo} J.A. Gupta et al., J. Cryst. Growth {\bf 231},
 48 (2001) and references therein.
\bibitem{molecN}
T. H. Myers et al.,
J.Vac.Sci.Technol. B {\bf 17}, 1654 (1999)
\end{thebibliography}
\end{document}